\documentclass[leqno,oneside]{article}
\usepackage{amsmath, amssymb}
\usepackage{amsfonts}
\newcommand{\field}[1]{\mathbb{#1}}
\newcommand{\bn}{$\boldmath$ \nabla $\unboldmath$}
\newcommand{\g}{\tilde{g}}
\newcommand{\Ti}{\tilde{T}}
\newcommand{\Kt}{\tilde{\hat{K}}}
\newcommand{\K}{\tilde{K}}
\newcommand{\N}{\tilde{N}}
\newcommand{\Rt}{\tilde{\hat{Ric}}}
\newcommand{\n}{\noindent}
\newcommand{\be}{\begin{equation}}
\newcommand{\ee}{\end{equation}}
\newcommand{\ben}{\begin{displaymath}}
\newcommand{\een}{\end{displaymath}}
\newcommand{\theone}{1\ }
\newcommand{\thetwo}{2\ }
\newcommand{\ep}{\hspace{\stretch{1}}$\Box$}
\usepackage{epsfig}
\usepackage{latexsym}
\usepackage{mathrsfs}
\newtheorem{R}{Remark}
\newtheorem{Def}{Definition}
\newtheorem{Prop}{Proposition}
\newtheorem{T}{Theorem}
\newtheorem{Con}{Conjecture}
\newtheorem{Lem}{Lemma}
\newtheorem{Cor}{\bf Corollary}
\begin{document}

\begin{center}
{\Large\bf On the asymptotic spectrum of the reduced volume in cosmological solutions of the Einstein equations}

\vspace{0.4cm}

{\large Martin Reiris}\footnote{e-mail: reiris@math.mit.edu.}\\

\vspace{0.1cm}

\textsc{Math. Dep. Massachusetts Institute of Technology}\\

\end{center}

\vspace{0.3cm}
\begin{center}
\begin{minipage}[c]{11cm}
\linespread{1.1}%
\selectfont
{\small Say $\Sigma$ is a compact three-manifold with non-positive Yamabe invariant. We prove that in any long time 
constant mean curvature Einstein flow over $\Sigma$, having bounded $C^{\alpha}$ space-time curvature at the cosmological scale, the reduced volume 
${\mathcal{V}}=(\frac{-k}{3})^{3}Vol_{g(k)}(\Sigma)$ 
($g(k)$ is the evolving spatial three-metric and $k$ the mean curvature) decays monotonically towards the volume 
value of the 
geometrization in which the cosmologically normalized flow decays. In more basic terms, under the given assumptions, 
there is volume collapse in the regions where the injectivity radius collapses (i.e. 
tends to zero) in the long time. We conjecture that under the curvature assumption above 
the Thurston geometrization is the unique global attractor.  We validate it in some special cases.}
 
\end{minipage}
\end{center}

\vspace{0.4cm}
\begin{center}
\section{Introduction}
\end{center}
\n A long-standing problem in General Relativity is to understand the long time evolution of cosmological 
solutions (solutions with compact space-like sections) of the Einstein equations at the cosmological scale or, in 
other words, to understand the large-scale shape of general cosmological solutions. Put in full mathematical 
generality the problem is outstandingly difficult and at present out of reach\footnote{The area of
cosmology, for which understand the large-scale shape is of central interest, overcame 
the mathematical difficulty by assuming large-scale homogeneity and isotropy and that only the averaged properties 
of matter contribute to the dynamic at the large-scales. The assumption reduces the mathematical complexity to the 
study of three well known models: the ${\mathcal{K}}=-1,0,1$ 
Friedman-Lema\^itre cosmologies. It is worth to remark that the justification of such assumption has now become a 
problem itself, the so called {\it averaging problem in cosmology}\cite{B2} which is the center of a large debate 
these days. The ${\mathcal{K}}=0,-1$
Friedman-Lema\^itre models have non-compact spatial sections, isometric to the flat $\field{R}^{3}$ for the ${\mathcal{K}}=0$ 
FL-model or the 
hyperbolic three-space
(with dynamical sectional curvature) still diffeomorphic to $\field{R}^{3}$ for the ${\mathcal{K}}=-1$ case. 
Both models however can be compactified
to obtain cosmological solutions (with compact slices). Perturbations 
around those models have also been largely studied by cosmologists.}. In this article we will present some progress 
in this problem for solutions satisfying suitable assumptions. More in particular we will 
investigate cosmological solutions of the Einstein constant mean curvature (CMC) flow equations over three-manifolds $\Sigma$ with 
non-positive Yamabe invariant (see later) and
having a uniform (in time) bound in the $C^{\alpha}$ space-time curvature (see later) at the cosmological scale
\footnote{The curvature assumption explicitly prohibits the formation of
singularities. In this sense the present work is about the evolution of cosmological solutions which
do not develop singularities at the cosmological scale. From a topological perspective, it deals with solutions whose
large-scale shape is driven by the topology of the three-dimensional space-like sections.}. 

Since stating the results with precision needs some technical elaboration, we will start by giving below a first glance of 
the ideas but in an informal manner. That may give a first flavor of the contents. After that we will comment on related developments 
and immediately thereafter we shall be introducing some primary terminology (as not all of it is standard in the field) and 
use it to give a detailed description of the contents in the rest of the article.

Consider a cosmological solution of the Einstein equations admitting a Cauchy hypersurface $\sim \Sigma$ of 
constant mean curvature (CMC) different from zero. Assume the Yamabe invariant\footnote{The Yamabe invariant of a 
compact three-manifold
is defined as the supremum of the scalar curvatures of unit volume Yamabe metric. Yamabe metrics are metrics
minimizing the Yamabe functional $(\int_{\Sigma}R_{g}dv_{g}/V_{g}^{\frac{1}{3}})$ over a fixed conformal class $[g]$. 
The Yamabe invariant is also known as {\it sigma constant} (see for instance \cite{FM2}).} $Y(\Sigma)$ of $\Sigma$ is non-positive. 
We will look at the flow $(g,K)$ along the (unique) 
CMC foliation where $g$ is the three-metric inherited from the space-time metric ${\bf g}$ and $K$ the second fundamental 
form at every CMC slice. The main object of study will be the {\it cosmologically normalized (CMC) flow}, namely the
flow $(\tilde{g},\tilde{K})=((\frac{k}{3})^{2}g,\frac{-k}{3}K)$ (see later). In elementary terms the main result will 
be to show that if the space-time curvature at the cosmological scale has uniformly (in time) bounded $C^{\alpha}$ norm 
with respect to every slice in the CMC foliation then as the mean curvature $k$ tends to zero\footnote{Note that when $k\rightarrow 0$ the cosmological time $t=-1/k$ diverges. We will use
the terminology ``{\it in the long time}'' to mean ``when $k\rightarrow 0$".}
 the flow $(\tilde{g},\tilde{K})$ separates the manifold $\Sigma$
persistently into a (possibly empty) $H$ (-hyperbolic) sector and a (possibly empty) $G$ (-graph) sector 
with particular properties that we describe next. The $H$ sector consists of a 
finite
set of manifolds admitting a complete hyperbolic metric $g_{H}$ of finite volume. Over each one of the $H$ pieces
the flow $(\tilde{g},\tilde{K})$ converges to $(g_{H},-g_{H})$ in the long time. The $G$ sector is instead a graph 
manifold\footnote{A graph manifold is a manifold obtained as a sum
along two-tori of $U(1)$ bundles over two-surfaces.} and over it
the injectivity radius collapses (i.e. tends to zero) at every point and in the long time. Moreover the 
volume of the $G$ sector
relative to the metric $\tilde{g}$ collapses to zero. This shows that in the long time, the volume of 
$\Sigma$ relative to the metric $\tilde{g}$, converges to the sum of the volumes of the hyperbolic pieces in the $H$ 
sector. The separation into the $H$ and $G$ sectors is called a {\it geometrization}. As we will explain later the results 
presented above
point towards a much deeper picture of the long time evolution of CMC solutions at the cosmological scale and under curvature 
bounds, namely
that the {\it Thurston geometrization} (see later) is the only global attractor.     

This article has its roots in the
works \cite{R}, \cite{FM}, \cite{A1}, \cite{FM2}. In \cite{FM2} Fischer and Moncrief studied for the first time 
the notion of volume collapse at the 
cosmological scale and its relations with the Yamabe invariant\footnote{The volume at the cosmological scale is the volume of 
$\Sigma$ relative to the metric $\tilde{g}$, namely ${\mathcal{V}}=(\frac{-k}{3})^{3}Vol_{g}$. We will call it either the {\it 
volume at the cosmological scale} or the {\it reduced volume} (see later). Note that Fischer and Moncrief use different 
terminology. They call {\it sigma constant} to what we call {\it Yamabe invariant} and 
{\it reduced Hamiltonian} to what we call {\it reduced volume}. We won't be following it here.}. In particular they 
investigated the reduced volume 
on a list of natural examples showing at least on those cases a connection between the asymptotic value of the reduced volume and the topology of the
Cauchy hypersurfaces. Their analysis validates the results of this article. A related 
investigation was carried out by Anderson in the seminal work \cite{A1}, where it is proved (also using the CMC gauge) that 
under 
pointwise curvature bounds (see the article for a precise statement) there is a sequence of CMC slices with $k\rightarrow 0$ 
on which the Einstein flow (suitable scaled) geometrizes the three-manifold. Similar
results but exploiting the reduced volume were obtained in \cite{R}. Finally, the notion of cosmological normalized flow 
that we use here was elaborated in \cite{R1} following \cite{AM}.

\vspace{2mm}
\n \begin{R} {\rm In the context of flows on manifolds with non positive Yamabe invariant, there are strong relations between the Einstein and the Ricci flow. In \cite{Ham} Hamilton
has been able to prove that under curvature bounds the Ricci flow geometrizes the manifold in much the same way as it has been proved here the Einstein flow does. 
He proves however that the tori separating the $H$ and $G$ pieces are incompressible and therefore the long time geometrization is the Thurston geometrization. It 
may be interesting to apply the results on volume collapse carried out in this paper to the Ricci flow under curvature bounds.}
\end{R}

We give next a more detailed description of the contents. In technical terms we will be dealing with space-times $(\bf{M},\bf{g})$ where $\bf{M}$ is a four-dimensional manifold
and $\bf{g}$ a $C^{\infty}$\footnote{We will assume all through that the Lorentzian space-time metric is of class $C^{\infty}$.} Lorentzian $(3,1)$ metric satisfying the Einstein equations in vacuum ${\bf Ric}=0$. 
Assume that there is a space-like slice of non-zero constant mean 
curvature (k) diffeomorphic to a three-dimensional manifold $\Sigma$. As is well known (see \cite{Re} and references therein) 
there is a unique region 
$\Omega_{CMC}$ inside 
${\bf M}$ and diffeomorphic to $\field{R}\times \Sigma$ where the mean curvature $k$ (which serves as a coordinate for 
the first factor) varies monotonically. Assume that $\Sigma$ is of non-positive Yamabe invariant $Y(\Sigma)$ (if 
$Y(\Sigma)>0$ it is conjectured \cite{Re}
that the flow becomes extinct in finite proper time in any of the two time-directions from any CMC slice and therefore the flow
would not be a long time flow). In this situation it is easy to check from the energy constraint that
$k$ never becomes zero. The existence of a CMC slice of non zero mean curvature defines two different 
time directions in $\Omega_{CMC}$: the direction in which the CMC slices increase volume that we will call ``the future" and the
direction in which they decrease volume that we will call ``the past"\footnote{By the Hawking singularity theorem all past directed time-like geodesics starting at a common CMC slice terminate before a uniform time lapse.}. We are interested in the dynamics in the future
direction. The CMC foliation induces a 3+1 splitting which allows us to write the metric $\bf{g}$ as
\begin{equation}\label{first}
{\bf g}=-(N^{2}-|X|^{2})dk^{2}+X^{*}\otimes dk+dk\otimes X^{*}+g,
\end{equation}

\n where $N$ is the {\it lapse function}, $X$ the {\it shift vector} and $g$ is a three-Riemannian metric on $\Sigma$ (depending on $k$). 
Thus the space-time metric $\bf{g}$ is described by a flow $(N,X,g)(k)$ that we will call the {\it Einstein (CMC) flow}. 
Let $T$ be the normal vector field to the CMC foliation and pointing in the future direction. The second fundamental form
$K$ of the CMC slices is $K=-\frac{1}{2}{\mathcal{L}}_{T}g$\footnote{${\mathcal{L}}_{X}$ denotes the Lie derivative along the vector field $X$}  and therefore $k=tr_{g}K$. The Einstein equations ${\bf Ric}=0$ 
in the CMC 3+1 splitting are  
\begin{equation}\label{constraints1}
R=|K|^{2}-k^{2},
\end{equation}
\begin{equation}\label{constraints2}
\nabla .K=0,
\end{equation}
\begin{equation}\label{h-j1}
\dot{g}=-2NK+{\mathcal{L}}_{X}g,
\end{equation}
\begin{equation}\label{h-j2}
\dot{K}=-\nabla\nabla N+N(Ric+kK-2K\circ K)+{\mathcal{L}}_{X}K,
\end{equation}
\begin{equation}\label{lapse}
-\Delta N+|K|^{2}N=1.
\end{equation}

\noindent Equations (\ref{constraints1}),(\ref{constraints2}) are the {\it constraint equations}, equations (\ref{h-j1}),(\ref{h-j2}) are the 
{\it Hamilton-Jacobi} 
equations of motion and (\ref{lapse}) is the (fundamental) 
{\it lapse equation} which is obtained after contraction of (\ref{h-j2}). Thus $N$ gets uniquely determined from
$(g,K)$ after solving (\ref{lapse}). Different choices of the shift vector give different flows $(X,N,g)$ over $\Sigma$ but 
the space-time solutions ${\bf g}$ they represent via equation (\ref{first}) are isometric. Thus up to space-time
diffeomorphism the Einstein flow is uniquely determined from the (abstract) flow $(g,K)(k)$. We will use the choice of $X=0$ 
all through the article. 

In cosmological terms the mean 
curvature $k$ is a measure of the universe expansion and can be identified \cite{R1} with $-3{\mathcal{H}}$ where ${\mathcal{H}}$ is the Hubble parameter 
(constant over each slice of the CMC foliation). At a slice $\{k_{0}\}\times \Sigma$ the Hubble parameter is 
${\mathcal{H}}_{0}=-\frac{k_{0}}{3}$ and if we scale
the space-time metric ${\bf g}$ as $\tilde{\bf g}={\mathcal{H}}_{0}^{2}{\bf g}$ we get a new space-time metric 
which is a new solution
of the Einstein equations in vacuum with three-metric $\tilde{g}={\mathcal{H}}_{0}^{2}g$ 
and second fundamental
form $\tilde{K}_{0}={\mathcal{H}}_{0}K$ at the same slice, thus having Hubble parameter equal to one (only in that 
slice). If  we perform such scaling at every
slice in the CMC foliation we obtain a flow $(\tilde{g},\tilde{K})({\mathcal{H}})=({\mathcal{H}}^{2}g,{\mathcal{H}}K)({\mathcal{H}})$ 
which we
will call the {\it cosmologically normalized Einstein flow} or {\it the Einstein flow at the cosmological scale}. The 
cosmologically normalized flow is the subject of the present article. Cosmologically normalized tensors will be denoted 
with a tilde either above or next to them. 
For example the space-time Riemannian tensor ${\bf Rm}_{\alpha\beta\gamma}^{\ \ \ \delta}$ is scale invariant, therefore
the cosmologically normalized Riemann curvature tensor is itself. The normal unit vector field $T$ scale as $T/{\mathcal{H}}$ 
and the combination $E={\bf Rm}_{\alpha\beta\gamma\delta}T^{\alpha}T^{\gamma}$ (the electric component of ${\bf Rm}$) is 
scale invariant. We will study the cosmologically normalized flow under the following {\it curvature assumption}.

\vspace{0.2cm}
\n {\bf Curvature assumption}\footnote{It is fundamental that we assume pointwise bounds of the curvature, i.e. 
bounds in the $C^{\alpha}$ norm of ${\bf Rm}$. $L^{2}$ bounds instead seem too weak to control the geometry in the 
thin parts.}: 
{\it there is a constant $\Lambda>0$ such that, at any time ${\mathcal{H}}$, the $C^{\alpha}_{\tilde{g}}({\mathcal{H}})$ norm of the cosmological normalized Riemann tensor $\tilde{{\bf Rm}}_{\alpha\beta\gamma}^{\ \ \ \delta}(={\bf Rm}_{\alpha\beta\gamma}^{\ \ \ \delta})$
 is bounded above by $\Lambda$}. 
\n \begin{R} ({\it on the $C^{\alpha}$ norm of the Riemann tensor}). {\rm Given a slice $\{{\mathcal{H}}\}\times \Sigma$ we decompose
the space-time Riemann tensor ${\bf Rm}$ into its electric $E_{\alpha\gamma}={\bf Rm}_{\alpha\beta\gamma\delta}T^{\beta}T^{\delta}$
and magnetic component $B_{\alpha\gamma}={\bf Rm}^{*}_{\alpha\beta\gamma\delta}T^{\beta}T^{\delta}$, where $^{*}$ means Hodge dual (see \cite{CK}). Now $E$ and $B$ are
two $(2,0)$, T-null tensors, which are symmetric and traceless. The $C^{\alpha}_{\tilde{g}}$ norm of ${\bf Rm}$ in the slice 
$\{{\mathcal{H}}\}\times \Sigma$ 
is defined as the 
$C^{\alpha}_{\g}$ norms of $E$ and $B$ as tensors in the Riemannian manifold $(\Sigma,\tilde{g}({\mathcal{H}}))$ 
(see the background section for a definition of the $C^{\alpha}_{\g}$ norm of a tensor). These $C^{\alpha}_{\g}$ 
norms are assumed to be uniformly bounded by $\Lambda$ for all ${\mathcal{H}}$ along the evolution.}  
\end{R}

\vspace{-0.3cm}
\n \begin{R} {\rm There is an example due to Ringstr\"om\footnote{The investigation is in connection with 
the a priori curvature condition given in \cite{A1}. It is easy to show that the example doesn't satisfy
our curvature assumption continuously in time.} \cite{Ring} (Prop 2) of a homogeneous Bianchi VIII model, 
showing that  while there are no singularities being formed the curvature assumption above is only satisfied over a 
divergent sequence of times, but not for all. The existence of such
a sequence is enough to apply many of the results of this article and to conclude in particular 
volume collapse.}
\end{R}  

The first main result will be the following.  

\begin{T}\label{theo1} Say $Y(\Sigma)\leq 0$ and say $(\tilde{g},\tilde{K})$ is a cosmologically normalized Einstein flow (in vacuum) satisfying the
curvature assumption. Then the range of ${\mathcal{H}}$ is of the form $(0,a)$ (it is a long time flow) and as 
${\mathcal{H}}\rightarrow 0$ (i.e. in the long time) the flow (weakly or strongly) persistently geometrizes the manifold $\Sigma$.
\end{T}

\begin{figure}[h]
\centering
\includegraphics[width=120mm,height=70mm]{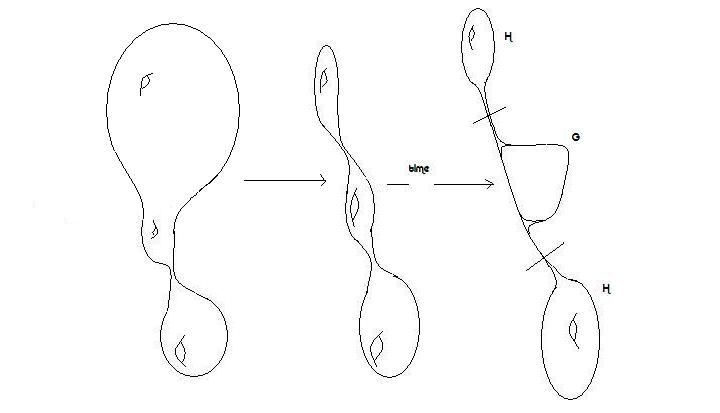}
\caption[U]{Large-scale picture of a cosmological solution. Observe that the reduced volume ${\mathcal{V}}$ is represented as
decreasing.}\label{fig1}
\end{figure}

\n Let us explain what a weak or strong geometrization is. Recall first the {\it thick-thin decomposition} of a Riemannian manifold\footnote{The Thick-thin decomposition
is a well know and standard separation of a Riemannian manifold.}.
Denote by $\Sigma^{\epsilon}$ the set of points in $(\Sigma,\tilde{g})$ where the injectivity radius is bounded below
by $\epsilon$ and $\Sigma_{\epsilon}$ the set of points where the injectivity radius is bounded above by $\epsilon$.
$\Sigma^{\epsilon}$ and $\Sigma_{\epsilon}$ are called the $\epsilon$-thick and $\epsilon$-thin parts of $(\Sigma,\tilde{g})$
and such decomposition is called the $\epsilon$-{\it thick-thin decomposition}. 
A flow $(\tilde{g},\tilde{K})$ in $\Sigma$ geometries $\Sigma$ iff there is (a continuous) $\epsilon({\mathcal{H}})$ with $\epsilon({\mathcal{H}})\rightarrow 0$
as ${\mathcal{H}}\rightarrow 0$ such that after a sufficiently
long time (i.e. after ${\mathcal{H}}$ gets sufficiently small) $\Sigma_{\epsilon}$ is persistently diffeomorphic
to a graph manifold to be denoted by $G$ and $\Sigma^{\epsilon}$ is persistently diffeomorphic to a finite set of
manifolds ($H_{i}$), to be denoted as $H$, admitting a complete hyperbolic metric of finite volume ($\g_{H,i}$) and with
$(\Sigma^{\epsilon},(\tilde{g},\tilde{K}))$ converging to $\cup_{i=1}^{i=n}(H_{i},(\g_{H,i},-\g_{H,i}))$ in $C^{2,\beta}\times C^{1,\beta}$ (see the background section for a precise description of the convergence). The manifolds separating the $G$ and $H$ sectors are
two tori. If all the tori are incompressible (their fundamental groups inject into the fundamental group of $\Sigma$) 
the geometrization is said to be {\it strong} and well known to be unique (see for instance \cite{R} Theorem 9), (actually equivalent to the Thurston decomposition
of the manifold). If one of the tori is not incompressible, the geometrization is said to be {\it weak}. A schematic picture 
of a geometrization is given in Figure \ref{fig1}. Let us exemplify the geometrization 
phenomenon with some simple but illustrative cases. 

\begin{enumerate}
\item $Y(\Sigma)<0$. The {\it flat cone or Robertson-Walker ${\mathcal{K}}=-1$} solution is ${\bf g}=-dt^{2}+t^{2}g_{H}$ where
$g_{H}$ is a hyperbolic metric on a hyperbolic manifold $\Sigma_{H}$. The mean
curvature is $k=\frac{-3}{t}$ and the normalized flow converges (it is actually
steady) to
$(g_{H},-g_{H})$ on the three dimensional manifold $\Sigma_{H}$. The solution
is flat.

\item $Y(\Sigma)=0$. 

\begin{enumerate} \item Consider now the solution ${\bf g}=-dt^{2}+\frac{t^{2}}{4}\sigma
+d\theta^{2}$ on $\Sigma=S_{gen}\times U(1)$, where $S_{gen}$ is a compact
surface of genus $gen>1$, $\sigma$ is a metric of constant scalar curvature equal to $-1$ 
on $S_{gen}$ and $d\theta^{2}$ is the standard element of length on $U(1)$. The mean curvature
is $k=\frac{-2}{t}$ and the normalized flow collapses to
a state $(\frac{\sigma}{9},-\frac{2\sigma}{3})$ on the two dimensional manifold
$S_{gen}$. The solution is flat.

\item The {\it Kasner} $(1,0,0)$ (with unit coefficients) is defined as ${\bf
g}=-dt^{2}+t^{2}d\theta_{1}^{2}+d\theta_{2}^{2}+d\theta_{3}^{2}$ on
$\Sigma=T^{3}$. The mean curvature $k=\frac{-1}{t}$ and the normalized flow
collapses to a state $(\frac{1}{9}d\theta^{2}_{1}$,
$\frac{-1}{3}d\theta_{1}^{2})$ on the one dimensional manifold $U(1)$. The
solution is flat.

\item The {\it Kasner} $(\frac{2}{3},\frac{2}{3},\frac{-1}{3})$ with unit coefficients is defined
as ${\bf g}=-dt^{2} +
t^{\frac{4}{3}}d\theta_{1}^{2}+t^{\frac{4}{3}}d\theta_{2}^{2}+t^{\frac{-2}{3}}d\theta_{3}^{2}$
on $\Sigma=T^{3}$. The mean curvature is $k=\frac{-1}{t}$ and the normalized
flow collapses with bounded curvature to a point, i.e. to the zero dimensional
space.  

\end{enumerate}
\end{enumerate}    

\n A crucial quantity used in the proof of Theorem \theone is the {\it reduced volume} ${\mathcal{V}}={\mathcal{H}}^{3}V_{g({\mathcal{H}})}$ which is the volume of the 
cosmologically normalized metric $\tilde{g}$\footnote{To our knowledge Fischer and Moncrief were the first to consider the reduced volume in the context of long time
evolution in the CMC gauge.}. As it turns out \cite{FM} the reduced volume (which is scale invariant) is 
either monotonically decreasing or
steady in which case the solution is a flat cone. Equally important, the infimum of the reduced volume when it is thought 
as a function on CMC states $(g,K)$ (i.e. pairs $(g,K)$ satisfying the constraint equations) is given by ${\mathcal{V}}_{inf}
=(\frac{-Y(\Sigma)}{-6})^{\frac{3}{2}}$ \cite{FM}.
The natural question is whether it is always the case that ${\mathcal{V}}\downarrow {\mathcal{V}}_{inf}$ at least 
under the curvature assumption above.
If it does so, it is known (\cite{R} Theorem 9) that the geometrization is strong and (therefore) unique. We will call it  the 
{\it Thurston geometrization}. We conjecture that such is always the case for solutions satisfying the curvature assumption.

\begin{Con}\label{con1} Say $Y(\Sigma)\leq 0$ and say $(\tilde{g},\tilde{K})$ is a cosmologically normalized Einstein flow satisfying the 
assumption. Then ${\mathcal{V}}\downarrow {\mathcal{V}}_{inf}=(\frac{-Y(\Sigma)}{6})^{\frac{3}{2}}$.
\end{Con}

\n Another way to express the conjecture is that the Thurston geometrization is a global attractor for cosmologically normalized 
flows satisfying the curvature assumption on manifolds with non-positive Yamabe invariant. If valid, the conjecture implies 
that the (scale invariant) Yamabe functional 
\ben
Y(g)=\frac{\int_{\Sigma}R_{g}dv_{g}}{V_{g}^{\frac{1}{3}}},
\een

\n converges to the Yamabe invariant $Y(\Sigma)$ along the flow. This can be sketchily seen as follows. Under the curvature assumption the scalar curvature
$R_{\tilde{g}}$ is known to be bounded (above and below, see Prop \ref{P1} later). On the other hand it is known that for manifolds with
$Y(\Sigma)\leq 0$ it is \cite{A3} $(\frac{-Y(\Sigma)}{6})^{\frac{3}{2}}=\sum_{i=1}^{i=n} Vol_{\g_{H,i}}(H_{i})$
where $H_{i}$ are the hyperbolic pieces in the Thurston decomposition of $\Sigma$. If ${\mathcal{V}}\downarrow {\mathcal{V}}_{inf}$
the volume of the $G$ sector collapses to zero and therefore $\frac{\int_{G}R_{\tilde{g}}dv_{\tilde{g}}}{V_{\tilde{g}}^{\frac{1}{3}}}\rightarrow 0$.
As $R_{\tilde{g}}\rightarrow -6$ on the $H$ sector we have $Y(\tilde{g})=Y(g)\rightarrow Y(\Sigma)$ as desired.

The second main result will be to show that always the $G$ sector collapses in (reduced) volume
\footnote{We note that this is a non-trivial statement. Consider the two-manifold $[a,b]\times S^{1}$ with the time dependent 
metric $g=t^{2}dx^{2}+\frac{1}{t^{2}}d\theta^{2}$. The volume is the same for all $t$ but $inj_{g}\rightarrow 0$.}. 

\begin{T} Say $Y({\Sigma})\leq 0$ and say $(\tilde{g},\tilde{K})$ is a cosmologically normalized flow 
satisfying the curvature assumption. Then the reduced volume of the 
total space converges towards the volume value of the long time geometrization.
\end{T}

\n The {\it volume value} of the geometrization (see also the background section) is

\n $\sum_{i=1}^{i=n} V_{g_{H,i}}(H_{i})$. This result is a first step to prove the conjecture above. In fact it 
validates the conjecture in some particular cases described in the Corollaries \ref{cor1}-\ref{cor4} which will be proved after the proof 
of Theorem \thetwo. 

\begin{Cor}\label{cor1} Say $Y(\Sigma)\leq 0$. Given $\Lambda$ there is $\epsilon$ such that for any cosmologically normalized
flow satisfying the curvature assumption (with the same $\Lambda$) and having 
${\mathcal{V}}-{\mathcal{V}}_{inf}\leq \epsilon$ at an initial time, it is ${\mathcal{V}}\downarrow {\mathcal{V}}_{inf}$
in the long time.
\end{Cor}

\n In basic terms, what Corollary \ref{cor1} says is that if we restrict to the set of solutions satisfying the
curvature assumption with a fixed $\Lambda$ then the Thurston geometrization is stable (in the class).

\begin{Cor}\label{cor2} Say $Y(\Sigma)\leq 0$ and say $(\tilde{g},\tilde{K})$ is a cosmologically normalized flow satisfying the
curvature assumption that is locally collapsing at every point, i.e. there is no $H$ sector in the long time
geometrization. Then $Y(\Sigma)=0$ and ${\mathcal{V}}\downarrow {\mathcal{V}}_{inf}=0$.
\end{Cor} 
       
Let ${\mathcal{V}}_{0}$ be the infimum of the volumes of all complete hyperbolic manifolds 
(with sectional curvature normalized to one), with or without cusps (this number is known to be positive \cite{Th}). Then we
have

\begin{Cor}\label{cor3} Say $Y(\Sigma)\leq 0$ and say $(\tilde{g},\tilde{K})$ is a cosmologically normalized flow satisfying the
curvature assumption. If at an initial time it is ${\mathcal{V}}<{\mathcal{V}}_{0}$ then 
${\mathcal{V}}\downarrow {\mathcal{V}}_{inf}=0$ in the long time.
\end{Cor}
 
\n Corollary \ref{cor3} says that if we restrict to the class of solutions satisfying the curvature assumption (with variable $\Lambda$), 
there is a threshold ${\mathcal{V}}_{0}$ for ${\mathcal{V}}$ at the initial time, below which the long time
gemetrization has only a $G$ sector and the reduced volume collapses to zero in the long time.

\begin{Cor}\label{cor4} Say $Y(\Sigma)\leq 0$ and say $(\tilde{g},\tilde{K})$ is a cosmologically normalized flow satisfying the
curvature assumption above. Then ${\mathcal{V}}\downarrow {\mathcal{V}}_{inf}$ iff the tori separating the
$H$ and $G$ sectors in the long time geometrization of the flow are incompressible in $\Sigma$.
\end{Cor}

\n Corollary \ref{cor4} shows that to prove Conjecture \ref{con1} is sufficient to prove the topological fact that
the two-tori separating the $H$ and $G$ sectors are incompressible.

Finally as an outcome of the proof of Theorems \theone and \thetwo we will be able to prove

\begin{Cor}\label{cor5} Say $Y(\Sigma)\leq 0$ and say $(\g,\K)$ is a cosmologically normalized flow satisfying the curvature 
assumption. Then the Bel-Robinson energy $Q_{0}({\bf Rm})$ is an $o({\mathcal{H}})$ (i.e. $\lim_{{\mathcal{H}}\rightarrow 0}Q_{0}/{\mathcal{H}}=0$). 
\end{Cor}

\begin{center}
\section{Background and terminology}
\end{center}

{\center \subsection{Convergence and collapse of Riemannian manifolds}}

\vspace{0.1cm}
We will make use the Cheeger-Gromov theory of convergence and collapse of Riemannian three-manifolds under curvature bounds. 
In particular we will use extensively the following result (\cite{A1} Prop 4 and 5. See also \cite{Pet})

\begin{Prop}\label{CH-G} Let $(\Sigma_{i},g_{i})$ be a sequence of Riemannian three-manifolds with or without boundary, with
uniformly bounded $C^{\alpha}_{g_{i}}$ Ricci curvature, i.e. $\|Ric_{g_{i}}\|_{C^{\alpha}_{g_{i}}}\leq \Lambda$. 
We have

\begin{enumerate}
\item\label{1} say $\{x_{i}\}$ is a sequence of points such that $dist(x_{i},\partial \Sigma_{i})\rightarrow \infty$ and 
$inj_{g_{i}}(x_{i})\geq inj_{0}$. Then one can extract a subsequence $(\Sigma_{i_{j}},x_{i_{j}},g_{i_{j}})$ converging
in $C^{2,\beta}$ to $(\Sigma_{\infty},x_{\infty},g_{\infty})$ ($\beta<\alpha$),
\item\label{2} suppose non of the $(\Sigma_{i}$ is diffeomorphic to a closed space form (a quotient of $S^{3}$). Then 
for any $\epsilon>0$ and $inj_{0}$ there is $r(\epsilon,\Lambda,inj_{0})$ with $r\rightarrow \infty$ as $\epsilon\rightarrow 0$, 
such that if $dist(x_{i},\partial \Sigma_{i})\geq r$ there is a finite cover of the ball $B(x_{i},r)$ with 
$inj_{g_{i}}(x_{i})\sim inj_{0}$.
\end{enumerate}
\end{Prop}

\n A number of remarks are in order.

\vspace{0.1cm}
i) If $U$ is any tensor field on a Riemannian manifold $(\Sigma,g)$ then
the $C^{k,\alpha}_{g}$ norm of $U$ is defined as 
\ben
\begin{split}
\|U\|_{C^{k,\alpha}_{g}}&=sup_{x\in\Sigma}\{|U|_{g}(x)+|\nabla U|_{g}(x)+ \dots +|\nabla^{k} U|_{g}(x)+\\
&\quad sup_{y\in \Sigma} \{\frac{|\nabla^{k}U(x)-\nabla^{k} U(y)|}{dist(x,y)^{\alpha}}\}\}.
\end{split}
\een

\n The difference $\nabla^{k} U(x)-\nabla^{k} U(y)$ is by parallel transport along any shortest geodesic joining $x$ with $y$.

ii) The convergence in item {\it 1} in the Proposition above is in the following sense: there is a sequence of submanifolds $\tilde{\Sigma}_{i}\subset \Sigma_{i}$ with 
$x_{i}\in \tilde{\Sigma_{i}}$ and $dist(x_{i},\partial \tilde{\Sigma_{i}})\rightarrow \infty$ 
and a sequence of diffeomorphisms (onto the image) $\varphi_{i}:\tilde{\Sigma}_{i}\rightarrow \Sigma_{i}$ such that
$\|\varphi_{i*}(g_{i})-g_{\infty}\|_{C^{2,\beta}_{g_{\infty}}}\rightarrow 0$.

c) A sequence of tensors $U_{i}$ in $(\Sigma_{i},x_{i},g_{i})$ converge in $C^{k,\beta}$ to $U_{\infty}$ in $(\Sigma_{\infty},x_{\infty},g_{\infty})$
if $\|\varphi_{i*}U_{i}-U_{\infty}\|_{C^{k,\beta}_{g_{\infty}}}\rightarrow 0$. 

d) One practical consequence of Proposition \ref{CH-G} is that when it comes to find interior elliptic estimates of certain 
elliptic operators on collapsed regions, we may well assume (because we can unwrap) that at the given point $x$ where one 
wants to extract the estimate there is a chart $(z_{1},z_{2},z_{3})$ of harmonic coordinates covering the ball $B(x,inj_{0})$
with $g_{ij}$ (the components of $g$ in the chart $\{z\}$) bounded in $C^{2,\alpha}_{\{z\}}$ (now the norms are standard 
H\"older norms on
the chart $\{z\}$) by a constant $C(inj_{0},\Lambda)$. We will be using this fact repeatedly all through the article.     

{\center \subsection{Electric-magnetic decomposition of the space-time curvature and some related formulae}}

In this section we introduce the electric/magnetic decomposition of the space-time curvature and useful formulae. Non of the
properties presented is given with a proof. The reader can consult the references \cite{CK},\cite{AM} for a detailed account 
on Weyl fields.  

Let $T$ be the normal unit vector field in the future direction and say ${\bf g}$ is a vacuum solution of the
Einstein equations. Then the electric and magnetic fields of ${\bf Rm}$ are defined by
\ben
E_{ab}={\bf Rm}_{acbd}T^{c}T^{d},\ \ B_{ab}=^{*}{\bf Rm}_{acbd}T^{c}T^{d}.
\een

\n The electric and magnetic fields are traceless and $T$-null $(2,0)$ vectors. In terms of $g$ and $K$ they have the
expressions
\ben
E=Ric+kK-K\circ K,\ \ B=-Curl K,
\een

\n where $Curl$ is the operator on symmetric $(2,0)$ tensors defined as 
$(Curl A)_{ab}=\frac{1}{2}(\epsilon_{a}^{\ cd}\nabla_{d}A_{cb}+\epsilon_{b}^{\ cd}\nabla_{d} A_{ca})$ ($\epsilon_{abc}$ is the volume form).
We also have
\ben
Div E=K\wedge B,\ \ Div B=-K\wedge B,
\een

\n where $(Div A)_{a}=\nabla^{b}A_{ba}$ is the divergence and $\wedge$ is the operation $(A\wedge C)_{a}=\epsilon_{a}^{\ bc}A_{b}^{\ d}C_{dc}$.
Dynamically (under zero shift) we have
\ben
\dot{E}=N Curl B-\nabla N\wedge B-\frac{5}{2}N (E\times K)-\frac{2}{3} N <E,K>g-\frac{1}{2}NkE,
\een
\ben
\dot{B}=-N Curl E+\nabla N\wedge E-\frac{5}{2}N (B\times K)-\frac{2}{3}N <B,K>k-\frac{1}{2}NkB.
\een
 
\n the dot meaning derivative with respect to $k$, $<.,.>$ is the inner product and $\times$ is the operation
\ben
(A\times C)_{ab}=\epsilon_{a}^{\ cd}\epsilon_{b}^{\ ef}A_{ce}C_{df}+\frac{1}{3}(A.C)g_{ab}-\frac{1}{3}(tr A)(tr C)g_{ab}.
\een 

The Bel-Robinson tensor is the totally symmetric, traceless, $(4,0)$ tensor $Q_{\alpha\beta\gamma\delta}$, defined
as
\ben
Q_{\alpha\beta\gamma\delta}={\bf Rm}_{\alpha\mu\gamma\nu}{\bf Rm}_{\beta\ \delta}^{\ \mu\ \nu}+{\bf Rm}_{\alpha\mu\gamma\nu}^{*}{\bf Rm}^{*\ \mu\ \nu}_
{\beta\ \delta}.
\een

\n We have $Q_{\alpha\beta\gamma\delta}({\bf Rm})T^{\alpha}T^{\beta}T^{\gamma}T^{\delta}=|E|^{2}_{g}+|B|^{2}_{g}$ and 
$\nabla^{\alpha}Q_{\alpha\beta\gamma\delta}({\bf Rm})=0$. Denote by $Q$ the integral in $\Sigma$ of $Q_{TTTT}$. Taking
the divergence of $Q_{\alpha TTT}$ and integrating we get the {\it Gauss equation}
\ben\label{GG}
\dot{Q}=-3\int_{\Sigma}NQ_{\alpha\beta TT}{\bf \Pi}^{\alpha\beta}dv_{g}.
\een

\n where ${\bf \Pi}$ is the {\it deformation tensor} ${\bf \Pi}_{\alpha\beta}=\bn_{\alpha}T_{\beta}$. Restricted to
any CMC slice we have ${\bf \Pi}=-K$, ${\bf \Pi}_{Ti}=\frac{1}{N}\nabla_{i}N$, ${\bf \Pi}_{Ti}=0$ and ${\bf \Pi}_{TT}=0$. 
Again dot means derivative with respect to $k$. All through the article we will use the formulae above but for the cosmologically 
normalized tensors. We will indicate that
by using a tilde above the referred tensor or by including a subindex $\g$ next to tensor operation, for instance $\wedge_{\g}$
or $| (.)|_{\g}$. The cosmological normalized versions of the equations above is straightforward to get and won't be deduced when needed.

\vspace{0.2cm}
{\center \subsection{The Newtonian potential, the reduced volume element and the logarithmic time}}

When working with cosmologically normalized quantities, it is convenient to use the {\it logarithmic time} $\sigma=-\ln -k$ as the time variable. 
Derivatives with respect to $\sigma$ of a cosmologically normalized quantity gives rise to a quantity which is also cosmologically normalized. 
{\it For the rest of the article derivatives with respect to $\sigma$ will be denoted with a dot.}

To illustrate how to work with cosmologically normalized quantities let us introduce the {\it Newtonian potential} 
$\phi=3\N-1=3N{\mathcal{H}}^{2}-1$ and
the {\it reduced volume element} $d\nu={\mathcal{H}}^{3}dv_{g}$ (both are scale invariant) and let us deduce the following 
pair of equations which will be fundamental
\be\label{locredvol}
\frac{\ln (d\nu)^{\frac{1}{3}}}{d\sigma}=\phi,
\ee
\be\label{Poisson}
\Delta_{\g}\phi-|\K|^{2}_{\g}\phi=|\Kt|^{2}_{\g}.
\ee

\n We have used a hat $\hat{}$ above $\K$ to mean the traceless part of $\K$ (with respect to $\g$)\footnote{We will use the same notation for $Ric$.}. 
Under zero shift we have 
\ben
\frac{d dv_{g}}{d k}=\frac{1}{2}tr_{g}\frac{d g}{d k}dv_{g}=-Nkdv_{g}.
\een

\n Now $d/d\sigma=(d/dk)(dk/d\sigma)=-kd/dk$ and so
\be
\frac{d \ln (d\nu)^{\frac{1}{3}}}{d\sigma}=\frac{d \ln {\mathcal{H}}}{d \sigma}+{\mathcal{H}}(N{\mathcal{H}}/3)=
3N{\mathcal{H}}^{2}-1,
\ee 

\n as desired. To get the Poisson-like equation for the Newtonian potential $\phi$ observe that the lapse equation (\ref{lapse}) is
scale invariant and therefore 
\ben
-\Delta_{\g}\N+|\K|^{2}_{\g}\N=1.
\een

\n Making $\N=\frac{1}{3}(\phi+1)$ we get
\ben
-\Delta_{\g}\phi+|\K|^{2}_{\g}\phi=3-|\K|^{2}_{\g}=-|\Kt|^{2}_{\g}.
\een

A final remark. The Maximum principle applied to (\ref{Poisson}) gives $-1\leq \phi\leq 0$. This important fact implies by
equation (\ref{locredvol}) that the local reduced volume element $d\nu$ is non increasing (under zero shift) and in 
particular that the reduced volume ${\mathcal{V}}$ is monotonically decreasing unless is steady in which case the solution 
is a flat cone. 

\vspace{0.2cm}
{\center\subsection{Geometric states and persistent geometric states}}

\n In this section we introduce some definitions, that although not strictly needed, puts the main concepts 
used in a broad geometric context.    

\begin{Def} Given a compact three-manifold $\Sigma$ define the geometric spectrum to be the set of all its
partitions (geometric states) of the form $\Sigma=\{H_{1}\ldots, H_{i},G_{1},$

\n $\ldots,G_{j}\}$ where the $H$ pieces 
are three manifolds
possibly with boundary admitting a complete hyperbolic metric of finite volume, and the $G$ are graph 
three-manifolds possibly with boundary. If any, the boundaries in all pieces are two-tori and a torus in the 
boundary of a $H$ piece is always a torus in the boundary of a $G$ piece. Two geometric states are said to
be equivalent if there is an isotopy in $\Sigma$ carrying the $H$ and the $G$ sectors of one into the $H$
and $G$ sectors of the other. A geometric state is said to be pure if there is only a $H$ or a $G$ piece.  
\end{Def}

\begin{Def} Given a geometric state $\{H_{1},\ldots,H_{i},G_{1},\ldots,G_{j}\}$, its volume value $V$ is defined
as the sum of the volumes of the complete hyperbolic metrics of finite volume of the $H$ pieces.
The volumetric spectrum is defined as the set of all the volume values for all the states in the geometric 
spectrum. 
\end{Def}

\n In this terminology, Theorem \theone says in particular that if $\epsilon$ is chosen sufficiently small
then after sufficiently long time the $\epsilon$-thick-thin decomposition is a (persistent) geometric state of the 
manifold. We will prove in Theorem \thetwo that ${\mathcal{V}}$ decreases to the 
volume value of the geometric state in which the flow is decaying. 

We make precise now the notion of persistence of a geometrization introduced in section $1$. These notions will be used in the
proofs of the main results. We say that a long time, cosmologically normalized flow $(\g,\K)$ implements a persistent geometrization iff either 

\begin{enumerate}

\item $inj_{\g(\sigma)}(\Sigma)\rightarrow 0$ as $\sigma$ goes to infinity (in which case there is only one persistent $G$ 
piece) or

\item $inj_{\g{\sigma}}(\Sigma)\geq inj_{0}>0$ as $\sigma$ goes to infinity (in which case there is only one 
persistent $H$ piece) and there is
a continuous function $\varphi:(-\ln -a,\infty)\times H\rightarrow \Sigma$, differentiable in the second factor, 
such that $\|\varphi ^{*}\g(\sigma)-\g_{H}\|_{C^{2,\beta}_{\g_{H}}}\rightarrow 0$ as $\sigma$ goes to infinity, or

\item the injectivity radius collapses in some regions and remains bounded below in some others (in which case 
there are a set of $G$ pieces $G_{1},\ldots,G_{j}$ and a set of $H$ pieces $H_{1},\ldots,H_{k}$) and for any 
$\epsilon>0$ and for
any $H$ piece $(H_{i},\g_{Hi})$ there is a continuous function $\varphi_{i}:(-\ln -a,\infty)\times 
H^{\epsilon}_{i}\rightarrow \Sigma$, differentiable in the second factor such that $\|\varphi_{i}^{*}\g(\sigma)-\g_{Hi}\|_{C^{2,\beta}_{\g_{Hi}}}
\rightarrow 0$ as $\sigma$ goes to infinity.

\end{enumerate}

{\center\subsection{Some useful terminology}}
 
Any sequence $\{\sigma_{i}\}$ of logarithmic times ($\sigma_{i}=-\ln -k_{i}$) which is diverging i.e. $\lim_{i\rightarrow \infty}\sigma_{i}=\infty$ 
will be called a {\it diverging sequence of logarithmic times} and abbreviated DSLT. Given a DSLT, $\{\sigma_{i}\}$, we say 
that a sequence of sets 
$\Omega(\sigma_{i})$ has {\it asymptotically total reduced volume} (ATRV) if 
${\mathcal{V}}(\Omega(\sigma_{i}))\rightarrow {\mathcal{V}}_{\infty}$
as $\sigma_{i}\rightarrow \infty$ where ${\mathcal{V}}_{\infty}$ is the limit of the reduced volume in the long time. 
Similarly we can define a set having {\it asymptotically non-zero} (ANZRV) or {\it asymptotically zero reduced volume} (AZRV). 
We say that a quantity $f$ controls a quantity $h$ if $|f|<M$ implies $|h|<C(M)$, and $f$ controls $h$ at zero if $M\rightarrow 0$ implies
$C(M)\rightarrow 0$.

\vspace{0.2cm}
{\center\section{Proof of the main results and corollaries}}

This section is organized as follows. We prove first three propositions (Propositions \ref{P1},\ref{P2},\ref{P3}) that would frame
the proofs of Theorem \theone and \thetwo. We prove then Theorems \theone and \thetwo and next the 
Corollaries \ref{cor1}-\ref{cor5}. 

Let us give a heuristic behind the proofs of Theorems \theone and \thetwo. The key ingredient is to look at
the reduced volume. As it is monotonic, it must settle in some limit value ${\mathcal{V}}_{\infty}$ as ${\mathcal{H}}\rightarrow 0$.
Therefore in the regions where the injectivity radius is bounded below (in $(\Sigma,\g(\sigma)))$ it must be $\phi\sim 0$
otherwise by equation (\ref{locredvol}) the volume would keep decreasing and eventually become below ${\mathcal{V}}_{\infty}$. 
Using equation (\ref{Poisson}) this implies $|\Kt|^{2}_{\g}\sim 0$ which after using the Einstein equations implies
$|\tilde{\hat{Ric}}|^{2}_{\g}\sim 0$. In other words the regions where the injectivity radius remains bounded below become 
hyperbolic. This argument gives in essence Theorem \theone. Theorem \thetwo is more involved because it deals with the regions
where the injectivity radius collapses. We are able to show however that if the $G$
regions (where the injectivity radius collapses) carry a non zero reduced volume (call it ${\mathcal{V}}_{0}$), 
then the regions inside $G$ whose unwrapped
geometry becomes hyperbolic carry asymptotically all the volume ${\mathcal{V}}_{0}$. This fact will imply an 
isoperimetric inequality showing that the regions lying at a 
distance between 1/2 and 1 from the collapsed regions and whose unwrapped geometry is becoming hyperbolic, carry also 
asymptotically a non zero reduced volume (if ${\mathcal{V}}_{0}\neq 0$). As these two regions are disjoint, the limit of the volume of the $G$ regions must be above ${\mathcal{V}}_{0}$
which is a contradiction.
    
\vspace{0.2cm}
\begin{Prop}\label{P1} Say $Y(\Sigma)\leq 0$ and say $(\tilde{g},\tilde{K})$ is a cosmologically normalized flow satisfying the
curvature assumption. At any logarithmic time $\sigma$ we have the following properties.

\begin{enumerate}

\item \label{i1} $\|\Kt\|_{C^{1,\alpha}_{\g}}$, $\|\Rt\|_{C^{\alpha}_{\g}}$ and $\|\phi\|_{C^{2,\alpha}_{\g}}$
are controlled by $\Lambda$.

\item For any $\epsilon>0$ there is $\delta(\epsilon,\Lambda)>0$ such that at any point $p$ if $|\phi(p)|\leq \delta$ then 
$|\Kt|_{\g}(p)\leq \epsilon$. In other words $-\phi(p)$ controls $|\Kt|_{\g}(p)$ at zero.

\end{enumerate}
\end{Prop}

\n {\bf Proof:} 

\n {\it 1}. As has been proved in \cite{A1} (Prop 2.2), $\|\K\|_{L^{\infty}_{\g}}$ and $\|Ric\|_{L^{\infty}_{\g}}$ are
controlled by $\Lambda$. Consider the elliptic system
\ben\label{eeo}
Div \Kt=0,
\een
\ben\label{eet}
Curl_{\g} \Kt=-B,
\een

\n Pick a point $x\in \Sigma$ and unwrap\footnote{So far we have $L^{\infty}_{\g}$ control of $\tilde{Ric}$. Still
proposition \ref{CH-G} holds, and one can unwrap to have $inj_{\g}\sim inj_{0}$ but this time the unwrapped geometry is controlled in
$C^{1,\beta}$. This is enough however to get elliptic estimates from equations \ref{eeo} and \ref{eet}. This is the only time we will need
an extension of proposition \ref{CH-G}.} if necessary to have $inj_{\g}(x)\geq inj_{0}$. Then interior Schauder
estimates \cite{Elliptic} show that $\|\Kt\|_{C^{1,\alpha}_{\g}(B(x,inj_{0}/2))}$ is controlled by $\Lambda$. Therefore $\|\Kt\|_{C^{1,\alpha}_{\g}}$ is
controlled by $\Lambda$. From $E=\tilde{Ric}-3\K+\K\circ\K$
we get that $\|\tilde{Ric}\|_{C^{\alpha}_{\g}}$ is controlled by $\Lambda$. Schauder estimates applied to
\ben
\Delta_{\g} \phi -|\Kt|^{2}_{\g}\phi=|\K|^{2}_{\g},
\een

\n show that $\|\phi\|_{C^{2,\alpha}_{\g}}$ is controlled by $\Lambda$.

\vspace{0.2cm}
\n {\it 2}. Suppose there is a sequence of logarithmic times $\{\sigma_{i}\}$ and a sequence of points $\{x_{i}\}$
such that $\phi(x_{i},\sigma_{i})\rightarrow 0$ but $|\Kt|_{\g(\sigma_{i})}(x_{i},\sigma_{i})\geq M$ with $M>0$. Unwrapping if 
necessary to have $inj_{\g}(x_{i})\geq inj_{0}$ we can extract a subsequence of $\{\sigma_{i}\}$ such that
on the balls $B(x_{i},inj_{0}/2)$, $\g(\sigma_{i})$ converges in $C^{2,\beta}$ to a limit metric $\g_{\infty}$, $\phi$ 
converges in $C^{2,\beta}$ to a limit
$\phi_{\infty}\leq 0$ with $\phi_{\infty}(x_{\infty})=0$ and $\Kt$ converges in $C^{1,\beta}$ to a limit $\Kt_{\infty}$ with $|\Kt|_{\g_{\infty}}
(x_{\infty})\geq M$, all satisfying the equation
\ben
\Delta_{\g_{\infty}}\phi_{\infty}-|\K_{\infty}|^{2}_{\g_{\infty}}\phi_{\infty}=|\Kt|^{2}_{\g_{\infty}}.
\een

\n However at $x_{\infty}$ it is $0\geq (\Delta_{\g_{\infty}}\phi_{\infty})(x_{\infty})=|\Kt|^{2}_{\g_{\infty}}(x_{\infty})>M>0$ which is
absurd. \hspace{\stretch{1}}$\Box$

\begin{Prop}\label{P2} Say $\Sigma$ is a compact three-manifold with 
bounded $C^{\alpha}_{g}$ norm of the curvature and bounded volume, i.e. $\|Ric\|_{C^{\alpha}_{g}}+Vol_{g}(\Sigma)\leq \Lambda$, 
and say $r<r'$. Then there is $C(\Lambda,r,r')$ such that for any measurable subset $\Omega$ it is $Vol_{g}(B(\Omega,r))\geq C(\Lambda,r,r') Vol_{g}(B(\Omega,r'))$
where $B(\Omega,s)$ is the ball of $\Omega$ with radius $s$. 
\end{Prop}

\n {\bf Proof:} 

Let ${\mathcal{K}}(\Lambda)<0$ be a lower bound for the sectional curvatures of any Riemannian three-manifold with 
$\|Ric\|_{L^{\infty}_{g}}\leq C(\Lambda)$.
Let $\{x_{i},i=1,\ldots,m\}\subset \Sigma$ be any set of $m$ points. By the Bishop-Gromov
volume comparison the function 

\n $Vol_{g}(\cup_{i=1}^{i=m}B(x_{i},r))/Vol_{g_{{\mathcal{K}}}}(o,r)$ is monotonically decreasing as $r$ increases, where 
$g_{\mathcal{K}}$ is a metric of constant sectional curvature ${\mathcal{K}}$ in $\field{R}^{3}$. We have therefore
\begin{equation}
Vol_{g}(\cup_{i=1}^{i=m}B(x_{i},r'))\leq C(\Lambda,r,r') Vol_{g}(\cup_{i=1}^{i=m}B(x_{i},r)),
\end{equation}

\n for any $r<\bar{r}$. Now consider a measurable set $\Omega$. There is $\{x_{i},i=1,\ldots,\infty\}\subset \Omega$ such that
\ben
\cup_{i=1}^{i=m}B(x_{i},r)\uparrow_{m} B(\Omega,r),
\een

\n and
\ben
\cup_{i=1}^{i=m}B(x_{i},r')\uparrow B(\Omega,r').
\een

\n Then taking volumes we have
\ben
\begin{split}
Vol_{g}(B(\Omega,r'))&=\lim_{m\rightarrow \infty} Vol_{g}(\cup_{i=1}^{i=m}B(x_{i},r'))\leq C(\Lambda,r,r')\lim_{m\rightarrow \infty} 
Vol_{g}(\cup_{i=1}^{i=m}B(x_{i},r))\\
&= C(\Lambda,r,r')Vol_{g}(B(\Omega,r)),
\end{split}
\een

\n which finishes the proof. \hspace{\stretch{1}}$\Box$\\

\begin{Prop}\label{P3} Say $Y(\Sigma)\leq 0$ and say $(\g,\K)$ is a cosmologically normalized flow satisfying the curvature assumption. We have the following
properties.

\begin{enumerate}
\item Given $\Gamma\geq 0$ and any DSLT, $\{\sigma_{i}\}$, the sequence of sets $\Omega_{\phi,\Gamma}(\sigma_{i})=\{
x\in \Sigma/ -\phi(x,\sigma_{i})\geq \Gamma\}$ has AZRV.

\item Given $\Gamma\geq 0$ and any DSLT, $\{\sigma_{i}\}$, the sequence of sets $\Omega_{\Kt,\Gamma}(\sigma_{i})=\{x\in\Sigma/|\Kt(x,\sigma_{i})|_{\g(\sigma_{i})}\geq \Gamma\}$
has AZRV.

\item Given $\Gamma\geq 0$ and any DSLT, $\{\sigma_{i}\}$, the sequence of sets $\Omega_{\nabla \Kt,\Gamma}=\{x\in \Sigma/|\nabla \Kt(x,\sigma_{i})|_{\g(\sigma_{i})}\geq \Gamma\}$
has AZRV.

\item For any pair of DSLT, $\{\sigma_{i}\}$ and $\{\sigma'_{i}\}$ with $\delta'\geq \sigma_{i}-\sigma'_{i}\geq \delta$ ($\delta'>\delta>0$ and fixed)
we have
\ben
\int_{\sigma'_{i}}^{\sigma_{i}}\|E\|^{2}_{L^{2}_{\g(\sigma)}}d\sigma \rightarrow 0.
\een

\item For any DSLT $\{\sigma_{i}\}$ we have
\ben
\tilde{Q}_{0}(\sigma_{i})=(\|E\|^{2}_{L^{2}_{\g}}+\|B\|^{2}_{L^{2}_{\g}})(\sigma_{i})\rightarrow 0.
\een

\n and therefore the sets 
\ben
\Omega_{B,\Gamma}(\sigma_{i})=\{x\in\Sigma/|B(x,\sigma_{i})|_{\g(\sigma_{i})}\geq \Gamma\},
\een
\ben
\Omega_{\tilde{\hat{Ric}},\Gamma}(\sigma_{i})=\{x\in \Sigma/|\tilde{\hat{Ric}}(x_{i},\sigma_{i})|_{\g(\sigma_{i})}\geq\Gamma\},
\een

\n have AZRV.

\end{enumerate}
\end{Prop} 

\n {\bf Proof:} 

\vspace{0.2cm}
\n {\it 1}. Differentiating 
\ben
\Delta_{\g}\phi-|\K|^{2}_{\g}\phi=|\Kt|^{2}_{\g},
\een

\n with respect to logarithmic time we get
\begin{equation}\label{Dder}
\Delta_{\g}\dot{\phi}-|\K|^{2}_{\g}\dot{\phi}=-({\Delta}_{\g})\dot{}\phi+(|\K|^{2}_{\g})\dot{}\phi+(|\Kt|^{2}_{\g})\dot{}.
\end{equation}

\n Appealing to
\ben
\dot{\g}=2\phi \g-6\tilde{N}\Kt,
\een
\ben
\dot{\Kt}=-\Kt-\phi\g-\nabla^{2}\phi+\phi E+E-\N(\Kt\circ\Kt-2\Kt),
\een
\ben
(\Delta_{\g})\dot{}(\phi)=<\nabla^{2}\phi,\dot{\g}>_{\g}-<\nabla\phi,Div \dot{\g}+\frac{1}{2}dtr_{\g}\dot{\g}>_{\g},
\een

\n we get by Proposition \ref{P1} {\it 1} that the right hand side of equation (\ref{Dder}) has $C^{\alpha}$ norm
controlled by $\Lambda$. By the maximum principle on equation (\ref{Dder}) $\|\dot{\phi}\|_{L^{\infty}}$ is controlled by $\Lambda$. Therefore
writing $\phi(x,\sigma)-\phi(x,\sigma_{i})=\int_{\sigma_{i}}^{\sigma}\dot{\phi}(x,\sigma)d\sigma$ we see that
if $-\phi(x,\sigma_{i})\geq \Gamma$ there is $T(\Lambda,\Gamma)$ such that $-\phi(x,\sigma)\geq \Gamma/2$ for every
$\sigma \in [\sigma_{i},\sigma_{i}+T(\Lambda,\Gamma)]$. Now suppose there is a subsequence of $\{\sigma_{i}\}$ denoted
by $\{\sigma_{i_{j}}\}$ such that ${\mathcal{V}}(\Omega_{\phi,\Gamma}(\sigma_{i_{j}}))\geq M$ for some $M>0$, then
$\dot{\mathcal{V}}(\sigma)=\int_{\Sigma}(d\nu)\dot{}=\int_{\Sigma}3\phi d\nu\leq -3MT(\Lambda)/2$ for any $\sigma\in
[\sigma_{i_{j}},\sigma_{i_{j}}+T(\Lambda,\Gamma)]$. Therefore as ${\mathcal{V}}$ is monotonic, 
it must be ${\mathcal{V}}(\sigma)\downarrow -\infty$ as $\sigma\rightarrow \infty$ which is absurd.

\vspace{0.2cm}
\n {\it 2}. This is direct from {\it 1} above and Proposition \ref{P1} {\it 2}.

\vspace{0.2cm}
\n {\it 3}. We prove first the claim that if $|\nabla \Kt(x,\sigma)|_{\g}\geq \Gamma$ then there is $r(\Lambda,\Gamma)$ and $x'\in B(x,r)$
such that $|\Kt(x',\sigma)|_{\g}\geq M$ for some $M(\Lambda,\Gamma)>0$. This shows that $\Omega_{\nabla\Kt,\Gamma}(\sigma)\subset B(\Omega_{\Kt,M}(\sigma),r)$.
Once this is proved, by Propositions \ref{P2} and \ref{P3} {\it 2} $B(\Omega_{\Kt,M}(\sigma_{i}),r)$ and therefore $\Omega_{\nabla\Kt,\Gamma}(\sigma_{i})$ have
AZRV which would finish this item. Now let us prove the claim. From now on if at $x$, $inj_{\g(\sigma)}(x)$ is small
we unwrap to have $inj_{\g(\sigma)}(x)\geq inj_{0}>0$. By Proposition \ref{P1} {\it 1} we have that $\nabla\Kt$ is controlled 
in $C^{\alpha}$, therefore
$|\nabla \Kt(x,\sigma)-\nabla \Kt(y,\sigma)|_{\g}\leq C(\Lambda) d(x,y)^{\alpha}$. Pick a unit vector $v(x)$ at $x$ such that
$|\nabla_{v}\Kt(x,\sigma)|_{\g}\geq \Gamma/3$. Pick a harmonic chart $\{x^{i}\}$ covering the ball $B(x,inj_{0})$
such that the Christoffel symbols $\Gamma_{ij}^{k}$ are zero at $x$ (this is always possible) and write in the $\{x^{i}\}$
coordinates
\ben
(\nabla_{v}\Kt)_{jk} (y)=v^{i}\partial_{x^{i}}\Kt_{jk}(y)-\Gamma_{jk'}^{l}(y)\Kt_{lk}(y)v^{k'}(y)-\Gamma_{kk'}^{l}(y)\Kt_{jl}(y)v^{k'}(y).
\een

\n Pick $r(\Lambda,\Gamma)\leq inj_{0}$ with $r^{\alpha}C(\Lambda)\leq \Gamma/20$ such that 
$|\Gamma_{ij}^{l}\Kt_{lk}|\leq \Gamma/40$ on $B(x,r)$. Then on $B(x,r)$ we have $|\partial_{v}\Kt(x)-\partial_{v}\Kt(y)|\leq \Gamma/10$
and therefore $|\Kt(p)-\Kt(q)|\geq \Gamma 2r/4$ where $p$ and $q$ are the intercepts of the line (in the coordinate
system $\{x^{i}\}$) $x+\lambda v$ and the boundary of the ball $B(x,r)$. So either $|\Kt(p)|$ or $|\Kt(q)|$ must be
greater or equal to $\Gamma r/4$. This finishes the proof of the claim.

\vspace{0.2cm} 
\n {\it 4}. We start by noting the following. Say $f$ and $h$ are tensorial quantities such that, given $\Gamma>0$ and any DSLT 
, $\{\sigma_{i}\}$, $\|h\|_{L^{\infty}_{\g}}(\sigma_{i})$ and $\|f\|_{L^{\infty}_{\tilde{g}}}$ are controlled by $\Lambda$ and 
the sequence of sets 
$\Omega_{f,\Gamma}(\sigma_{i})=\{x\in\Sigma/|f(x,\sigma_{i})|_{\g}\geq \Gamma\}$ has AZRV, then: 
a) for any DSLT, $\{\sigma_{i}\}$, it is $(\int_{\Sigma}|h*f|_{\g}^{2}dv_{\g})(\sigma_{i})\rightarrow 0$ ($*$ is some tensorial composition) 
and b) for any pair of DSLT, $\{\sigma_{i}\}$ and $\{\sigma'_{i}\}$ as
in the statement of this item (Prop \ref{P3}, {\it 4}), it is $\int_{\sigma'_{i}}^{\sigma_{i}}(\int_{\Sigma}|h*f|^{2}dv_{\g})d\sigma\rightarrow 0$ as 
$\sigma_{i}\rightarrow \infty$. The claim
a) is obvious by writing $|h*f|_{\g}\leq c|h|_{\g}|f|_{\g}$ for some numeric $c$. For the claim b) observe that if the claim holds
for $h=1$ it holds for any $h$. Now if it is false for $h=1$ we can extract a sequence of logarithmic times $\{\bar{\sigma}_{i}\}$
with $\sigma_{i}\geq \bar{\sigma}_{i}\geq \sigma'_{i}$ and $(\int_{\Sigma}|f|_{\g}^{2}dv_{\g})(\bar{\sigma_{i}})\nrightarrow 0$ which contradicts a).

Now note that by {\it 2} above, the claim a) and 
\ben
Curl_{\g}\K=-B,
\een

\n we have that for any DSLT $\{\bar{\sigma}_{i}\}$, $\|B\|_{L^{2}_{\g}}(\sigma_{i})\rightarrow 0$. 
     
We will prove this item by studying the quantity
\begin{equation}\label{form}
\int_{\Sigma}<E,\Kt>_{\g}dv_{\g},
\end{equation}

\n and its derivative with respect to logarithmic time. Differentiating it with respect to $\sigma$ we get
\begin{equation}\label{EK}
\begin{split}
(\int_{\Sigma}<E,\Kt>_{\g}dv_{\g})\dot{}&=\int_{\sigma} <\dot{E},\Kt>_{\g}+<E,\dot{\Kt}>_{\g}-<E\circ\Kt_{\g},\dot{\g}>+\\
&\quad +3<E,\Kt>_{\g}\phi dv_{\g}.
\end{split}
\end{equation}

\n To estimate the terms on the right hand side of the last equation we appeal to the equations
\begin{equation}\label{gdot}
\dot{\g}=2\phi \g-6\tilde{N}\Kt,
\end{equation}
\begin{equation}\label{Edot}
\dot{E}=\N Curl_{\g} B -\frac{\nabla \N}{\N}\wedge_{\g} B-\frac{5}{2}E\times_{\g} \K-\frac{2}{3}<E,\K>_{\g}\g-\frac{3}{2}E,
\end{equation}
\begin{equation}\label{Kdot}
\dot{\Kt}=-\Kt-\phi\g-\nabla^{2}\phi+\phi E+E-\N(\Kt\circ\Kt-2\Kt).
\end{equation}

\n Recall $3\N=\phi+1$. Integrate both sides of equation \ref{EK} in the interval $[\sigma'_{i},\sigma_{i}]$ and plug in
the equations (\ref{Edot}), (\ref{Kdot}), (\ref{gdot}). By the claim a) above the left hand side (of the integrated equation) 
converges to zero as $i\rightarrow \infty$. Using the items {\it 1}, {\it 2} and the claim b) above we get that
the only terms on the right hand side (of the integrated equation) that may not converge to zero as $i\rightarrow \infty$ are
\be\label{Ephi}
\int_{\sigma'_{i}}^{\sigma_{i}}\int_{\Sigma}<E,\nabla^{2}\phi>_{\g}dv_{\g}d\sigma,
\ee
\begin{equation}\label{BKt}
\int_{\sigma'_{i}}^{\sigma_{i}}\int_{\Sigma}\N <Curl_{\g} B,\Kt>_{g}dv_{\g}d\sigma,
\end{equation}

\n and
\begin{equation}\label{E^2}
\int_{\sigma'_{i}}^{\sigma_{i}}\int_{\Sigma}|E|^{2}_{\g}dv_{\g}d\sigma.
\end{equation}

\n To show that the equation (\ref{Ephi}) and (\ref{BKt}) converge to zero as $i\rightarrow \infty$ we integrate by parts.
In equation (\ref{Ephi}) integration by parts gives
\ben
\int_{\sigma'_{i}}^{\sigma_{i}}\int_{\Sigma}-<Div E,\nabla \phi>_{g}dv_{\g}d\sigma,
\een

\n which by the formula $Div_{\g} E=\K\wedge_{\g} B$ and the claim b) above is guaranteed to converge to zero as
$i\rightarrow \infty$. To integrate by parts on equation \ref{BKt} invoke the formula
\ben
Div(U\wedge U')=-<Curl U,U'>+<U,Curl U'>,
\een

\n holding for any $U$ and $U'$ traceless symmetric tensors. This gives 
\begin{equation}\label{intparts}
\int_{\sigma'_{i}}^{\sigma_{i}}\int_{\Sigma}<B,\nabla \N * \Kt+\N Curl\Kt>dv_{\g}d\sigma,
\end{equation}

\n ($*$ is some tensor operation). After using the formula $Curl_{\g}\Kt=-B$ in equation (\ref{intparts}), claim b)
above shows that all the expression goes to zero as $i\rightarrow \infty$. We are thus lead to conclude that the
expression (\ref{E^2}) goes to zero as $i\rightarrow \infty$ as desired.

\vspace{0.2cm}
\n {\it 5}. We have shown in ${\it 4}$ that $\|B\|^{2}_{L^{2}_{\g(\sigma_{i})}}$ goes to zero as $i\rightarrow \infty$
for any DSLT $\{\sigma_{i}\}$. To show that the same happens for $E$ we make use of the Bel-Robinson energy. Observe
that $\frac{1}{\mathcal{H}}\int_{\Sigma}(|E|_{g}^{2}+|B|^{2}_{g})dv_{g}
=\frac{1}{\mathcal{H}}\int_{\Sigma}Q_{TTTT}({\bf Rm})dv_{g}=\int_{\Sigma}Q_{\Ti\Ti\Ti\Ti}(\tilde{\bf Rm})dv_{\g}=
\int_{\Sigma}(|E|^{2}_{\g}+|B|^{2}_{\g})dv_{\g}$. Using the Gauss equation we get 
\begin{equation}\label{Gaussint}
\dot{\tilde{Q}}=\tilde{Q}-9\int_{\Sigma}\N \tilde{Q}_{\alpha\beta \Ti\Ti}\tilde{\bf \Pi}^{\alpha\beta}dv_{\g}.
\end{equation}

\n Now suppose there is a DSLT, $\{\sigma_{i}\}$, such that $\int_{\Sigma}|E|^{2}_{\g}dv_{g}>M$ for some $M>0$. As the
right hand side of equation (\ref{Gaussint}) is controlled by $\Lambda$ we can find $T(\Lambda, M)$ such that the 
integral in $\sigma$ of the right hand side of equation (\ref{Gaussint}) on the interval $[\sigma_{i},\sigma_{i}+T]$
is, in absolute value, less than $M/2$ and therefore integrating equation (\ref{Gaussint}) on the interval $[\sigma_{i},\sigma]$
with $\sigma\in [\sigma_{i},\sigma_{i}+T]$ we get $\tilde{Q}(\sigma)\geq M/2$. Therefore $\int_{\sigma_{i}}^{\sigma_{i}+T}\tilde{Q}d\sigma\geq TM/2$.
However by ${\it 4}$ we know $\int_{\sigma_{i}}^{\sigma_{i}+T}\int_{\Sigma}|E|^{2}_{\g}+|B|^{2}_{\g}dv_{\g}d\sigma\rightarrow
0$ which is a contradiction.     

\hspace{\stretch{1}}$\Box$\\

We are now ready to prove Theorems $1$ and $2$ (stated conveniently).

\begin{T} Say $Y(\Sigma)\leq 0$ and say $(\g,\K)$ a cosmologically normalized flow satisfying the curvature assumption. 
Then $(\g,\K)$ induces a unique persistent geometrization on $\Sigma$.
\end{T}

\n {\bf Proof:} 

We prove first there is a DSLT, $\{\sigma_{i}\}$ with
$(\Sigma^{\frac{1}{i}},(\g,\K)(\sigma_{i}))$ converging to 

\n $\cup_{i=1}^{i=n}(H_{i},(\g_{H,i},-\g_{H,i}))$ (in $C^{2,\beta}$).
Introduce a new variable $j=1,2,3,\ldots$. For $j=1$ find a sequence $\{\sigma_{1,i}\}$ with 
$(\Sigma^{1},\g(\sigma_{1,i}))$ convergent in $C^{2,\beta}$. For $j=2$ find a subsequence $\{\sigma_{2,i}\}$ 
of $\{\sigma_{1,i}\}$ with $(\Sigma^{1/2},\g(\sigma_{2,i}))$ convergent in $C^{2,\beta}$. Proceed similarly
for all $j$ to have a double sequence $\{\sigma_{j,i}\}$. Now, the diagonal sequence $\{\sigma_{i,i}\}$, $(\Sigma^{1/i},\g(\sigma_{i,i}))$
converges into a union of complete manifolds of finite volume, denoted as $\cup_{\nu} (M_{\nu},\g_{\infty,\nu})$. 
By Proposition \ref{P3} {\it 2}, $\K(\sigma_{i,i})$ converges to $-\g_{\infty,\nu}$ in $C^{1,\beta}$. By proposition \ref{P3} {\it 5} and
the formula $E=\tilde{Ric}-3\K+\K\circ\K$ we get that each metric $\g_{\infty,\nu}$ is hyperbolic. Therefore, as there
is a lower bound for the volume of complete hyperbolic manifolds of finite volume and the total volume of the limit 
space is bounded above, there must be a finite number of components, and we can write $\cup_{\nu} (M_{\nu},\g_{\infty,\nu})=\cup_{i=1}^{i=n}(H_{i},\g_{H,i})$.
 
We prove next that each component $(H_{j},\g_{H,j})$ is persistent. For simplicity assume there is only one component
and therefore $(\Sigma^{1/i},\g(\sigma_{i,i}))$ converges in $C^{2,\beta}$ to $(H,\g_{H})$. There are two possibilities according to whether
the component is compact ot not, we discuss them separately.

\begin{figure}[h]
\centering
\includegraphics[width=110mm,height=60mm]{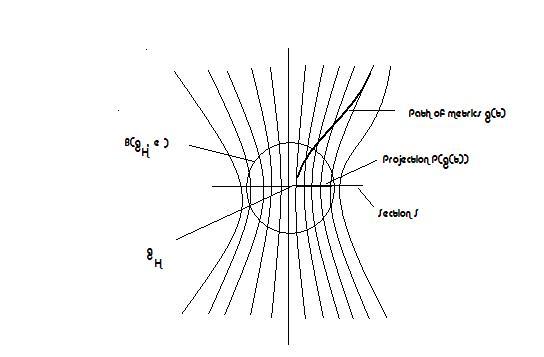}
\caption[U]{Representation of the space of metrics on a neighborhood of $\g_{H}$.}\label{fig2}
\end{figure}

1.{\it (The compact case)} Assume $(H,\g_{H})$ is compact. Consider the space of metrics ${\mathcal{M}}_{H}$ in $H$. For every metric $g$ consider
the orbit of $g$ under the diffeomorphism group. Denote such orbit by $o(g)$. Around $\g_{H}$
consider a small (smooth) section ${\mathcal{S}}$ of ${\mathcal{M}}_{H}$ of $C^{2,\beta}_{\g_{H}}$ metrics transverse to the 
orbits generated by the action on ${\mathcal{M}}_{H}$ of the diffeomorphism group
\footnote{Which particular section is taken is unimportant. One can use for instance ${\mathcal{S}}=\{g/id:(H,g)\rightarrow (H,\g_{H})\}$
is harmonic (see \cite{AM}, \cite{Ham}). The same comment applies in the non-compact case.}. For an illustration see Figure \ref{fig2}. 

If $\epsilon_{0}$ is sufficiently small 
every ($C^{\infty}$) metric $g$ in ${\mathcal{M}}_{H}$ with $\|g-\g_{H}\|_{C^{2,\beta}_{\g_{H}}}\leq \epsilon_{0}$ can be uniquely projected 
into ${\mathcal{S}}$
by a diffeomorphism, or in other words we can consider the projection $P(g)=o(g)\cap {\mathcal{S}}$. Note that one can project
every $(C^{\infty}$) path of ($C^{\infty}$) metrics $g(t)$ starting close to $\g_{H}$, to a $(C^{\infty})$ path $P(g(t))$, 
until at least the first time when $\|P(g(t))-\g_{H}\|_{C^{2,\beta}_{\g_{H}}}= \epsilon_{0}$
or in other words until at least when the projection touches the boundary of the ball of center $\g_{H}$ and radius $\epsilon_{0}$ 
in $C^{2,\beta}_{\g_{H}}$ (denote such ball as
$B(\g_{H},\epsilon_{0})$)\footnote{We consider $C^{\infty}$ paths of $C^{\infty}$ metrics because we have assumed the solution ${\bf g}$ and
therefore the zero shift flow to be $C^{\infty}$. It is not difficult to show that independent of the section ${\mathcal{S}}$ 
considered, a path $g(t)$ as above (leaving or not the ball $B(\g_{H},\epsilon)$) can be projected into ${\mathcal{S}}$ until at least a first time
when the projection touches the boundary of the ball. Note that if $\phi^{*}_{t}(g(t))=P(g(t))\in B(g_{H},\epsilon_{0})\cap {\mathcal{S}}$ for $t<t_{*}$ 
then for every $t_{1}<t_{*}$ $\phi_{t_{1}}^{*}(g(t))$ is a path in $B(g_{H},\epsilon_{0})$ for $t$ in a neighborhood of $t_{1}$ and which therefore can
be projected into ${\mathcal{S}}$.} 

Recall Mostow rigidity\footnote{Mostow rigidity says that any two hyperbolic metrics on a compact manifold are necesarily isometric. What we 
state as {\it Mostow rigidity} here is an obvious consequence of this fact.}
\vspace{0.1cm}
\n {\it Mostow rigidity (the compact case)}. There is $\epsilon_{1}$ such that if $P(g'_{H})\in B(\g_{H},\epsilon_{1})$, where
$g'_{H}$ is a hyperbolic metric in $H$ then $P(g'_{H})=\g_{H}$. 

\vspace{0.1cm}
Fix $\epsilon_{2}=min\{\epsilon_{0},\epsilon_{1}\}$. Observe that as $\g_{\sigma_{i,i}}\rightarrow \g_{H}$ in $C^{2,\beta}$ there is 
a sequence of diffeomorphisms $\phi_{i}$ such that $\phi^{*}_{i}(g(\sigma_{i,i}))$ converges to $\g_{H}$ in $C^{2,\beta}_{\g_{H}}$.
Now, if the geometrization is not persistent there is $\epsilon\leq \epsilon_{2}$ and $i_{2}$ such that if $i\geq i_{2}$ then 
$P(\phi^{*}_{i}(\g(\sigma)))$ is well defined for $\sigma\geq \sigma_{i,i}$ until a first time $\sigma_{i,i}+T_{i}$ when 
$P(\phi^{*}_{i}(\g(\sigma_{i,i}+T_{i})))$ is in $\partial B(\g_{H},\epsilon_{2})$. But we know the sequence of Riemannian
manifolds $(H,P(\phi^{*}_{i}(\g(\sigma_{i,i}+T_{i}))))$ converge in $C^{2,\beta}$ to $\g_{H}$, and that means by the definition of $C^{2,\beta}$ convergence and 
Mostow rigidity 
that there is a sequence of diffeomorphisms
$\varphi_{i}$ such that $P(\varphi^{*}_{i}(P(\phi^{*}_{i}(\g(\sigma_{i,i}+T_{i}))))$ converge to $\g_{H}$ in 
$C^{2,\beta}_{\g_{H}}$. This contradict the fact that $P(\phi^{*}_{i}(\g(\sigma_{i,i}+T_{i})))$ is in $\partial B(\g_{H},\epsilon_{2})$.

2. {\it (The non-compact case)}. The proof 
of this case proceeds along the same lines as the compact case but special care must be taken at the cusps. Let us assume for 
simplicity that there is only one cusp in the piece $(H,\g_{H})$. Given $A$
sufficiently small there is a unique torus transverse to the cusp, to be denoted by $T^{2}_{A}$, of constant mean curvature 
and area $A$. Denote by $H_{A}$ the ``bulk" side of the torus $T^{2}_{A}$ in $H$. Consider the set of metrics ${\mathcal{M}}_{H_{A}}$
on $H_{A}$ such that for any of them $T^{2}_{A}$ has constant mean curvature and area $A$. Consider the action of the diffeomorphism
group on ${\mathcal{M}}_{H_{A}}$ leaving the torus $T^{2}_{A}$ invariant. Again the orbit of $g$ will be denoted by $o(g)$.
Consider a small (smooth) section of ${\mathcal{S}}$ of $C^{2,\beta}$ metrics around $\g_{H}$ and transverse to the orbits of 
the action by the diffeomorphism group mentioned above. Finally consider the projection $P(g)=o(g)\cap {\mathcal{S}}$ which is
well defined on a ball $B(\g_{H},\epsilon_{0})$ for $\epsilon_{0}$ small enough. Observe again that a path $g(t)$ of
metrics in ${\mathcal{M}}_{H_{A}}$ can be projected into ${\mathcal{S}}$ until at least the first time 
when $P(g(t))$ is in $\partial B(\g_{H},\epsilon_{0})$. Slightly abusing the notation (as we would need a pointed sequence)
consider the sequence $(\Sigma,\g(\sigma_{i,i}))$ converging in $C^{2,\beta}$ to $\g_{H}$. If $i\geq i_{0}$ we can identify on $\Sigma$
a torus $T^{2}_{A,\g(\sigma_{i,i})}$ of constant mean curvature and area $A$ which converges as $i\rightarrow \infty$ (and after the application of a diffeomorphism) 
to $T^{2}_{A}$ in $H$\footnote{For a proof of this fact see the footnote in page 328 in \cite{Ham}.}. 
More in particular there is a sequence of diffeomorphisms (onto the image) 
$\phi_{i}:H_{A}\rightarrow \Sigma$ such that 
$\phi_{i}(T^{2}_{A})=T^{2}_{A,\g(\sigma_{i,i})}$ and $\|\phi^{*}_{i}(\g(\sigma_{i,i}))-\g_{H}\|_{C^{2,\beta}_{\g_{H}}}$
converging to zero. We note the following crucial facts (justified below).

i) The diffeomorphisms (onto the image) $\phi_{\sigma}:H_{A}\rightarrow \Sigma$ with $\phi_{\sigma}(T^{2}_{A})=T^{2}_{A,\g(\sigma)}$ can
be defined (varying differentiably) as long as the tori $T^{2}_{A,\g(\sigma)}$ are well defined (varying differentiably).

ii) There are $\sigma_{0}$ and $\epsilon_{1}$ such that if $\sigma_{1}\geq \sigma_{0}$ and $\phi_{\sigma_{1}}:H_{A}\rightarrow \Sigma$ is well 
defined and $\|\phi^{*}_{\sigma_{1}}(\g(\sigma_{1}))-\g_{H}\|_{C^{2,\beta}_{\g_{H}}}\leq \epsilon_{1}$ then the tori $T^{2}_{A,\g(\sigma)}$ are
well defined, varying differentiably for $\sigma$ on a neighborhood of $\sigma_{1}$.

iii) If for $\sigma_{1}\geq \sigma_{0}$ $\phi_{\sigma_{1}}:H_{A}\rightarrow \Sigma$ is well defined and satisfies 
$\|\phi^{*}_{\sigma_{1}}(\g(\sigma_{1}))-\g_{H}\|_{C^{2,\beta}_{\g_{H}}}\leq \epsilon_{1}$ then $\phi_{\sigma}$ is well defined 
until at least the first time $\sigma_{2}\geq \sigma_{1}$ when $\|P(\phi^{*}_{\sigma_{2}}(\g(\sigma_{2})))-\g_{H}\|_{C^{2,\beta}_{\g_{H}}}
=\epsilon_{1}$.
    
The fact i) is self evident. The fact ii) is the most important to consider and can be justified as follows. It is well known
that under curvature bounds the injectivity radius cannot collapse in finite distance from a region that is non collapsed. In
particular if $\|\phi^{*}_{\sigma_{1}}(\g(\sigma_{1}))-\g_{H}\|_{C^{2,\beta}_{\g_{H}}}\leq \epsilon_{1}$ for $\epsilon_{1}$ sufficiently small
the ``bulk" side of $T^{2}_{A,\g(\sigma_{1})}$ is non collapsed and therefore the `` cusp" side of $T^{2}_{A,\g(\sigma_{1}))}$
is non collapsed in finite distances from the ``bulk" side. Now, if $\sigma_{1}$ is big enough the $C^{\beta}$ norm of $\hat{Ric}$ around 
$T^{2}_{A,g(\sigma_{1})}$ must be small otherwise one may find a DSLT for which the pointed sequence 
$(\Sigma,p_{i},\g(\sigma_{i}))$ with $p_{i}\in T^{2}_{A,\g(\sigma_{i})}$
is not converging into a complete hyperbolic metric of finite volume. This shows in particular that if $\sigma_{0}$ is big 
enough the geometry nearby $T^{2}_{A,\g(\sigma_{1})}$ is close (in $C^{2,\beta}$) to the geometry nearby $T^{2}_{A}$ in $H$. 
By the continuity of the flow the tori $T^{2}_{A,\g(\sigma)}$ are well defined for $\sigma$ in a neighborhood of $\sigma_{1}$.
The fact iii) follows directly from ii).

Recall Mostow Rigidity\footnote{For a proof of this fact as well as for realted discussions the reader can consult \cite{Ham} (footnote on page 323).}

\vspace{0.1cm}
\n {\it Mostow rigidity (the non-compact case)}. There is $A_{0}$ such that for any $A\leq A_{0}$ there is $\epsilon_{0}$ such
that if $(\Sigma',g'_{H})$ is a complete hyperbolic manifold of finite volume and $\phi:H_{A}\rightarrow \Sigma'$ is a
diffeomorphism onto the image satisfying $\|\phi^{*}(g_{H}')-\g_{H}\|_{C^{2,\beta}_{\g_{H}}}\leq \epsilon_{0}$ then $(\Sigma',g'_{H})$
is isometric to $(H,\g_{H})$. In particular $P(\phi^{*}(g'_{H}))=\g_{H}$.

\vspace{0.1cm}

Given $A\leq A_{0}$ but so far arbitrary, fix $\epsilon_{2}=min\{\epsilon_{0},\epsilon_{1}\}$. Due to the facts i), ii) and iii) we have that if the 
geometrization is not persistent there is $\epsilon\leq \epsilon_{2}$ and $i_{2}$ such that if $i\geq i_{2}$ then 
$P(\phi^{*}_{i}(\g(\sigma)))$ is well defined for $\sigma\geq \sigma_{i,i}$ until a first time $\sigma_{i,i}+T_{i}$ when 
$P(\phi^{*}_{i}(\g(\sigma_{i,i}+T_{i})))$ is in $\partial B(\g_{H},\epsilon_{2})$. Now the sequence $P(\phi^{*}_{i}(\g(\sigma_{i,i}+T_{i})))$ 
has a subsequence converging in $C^{2,\beta}$ to a complete hyperbolic metric in finite volume. Again as in the compact case, by Mostow rigidity it must be
converging in $C^{2,\beta}_{\g_{H}}$ to $\g_{H}$ contradicting the fact that $P(\phi^{*}_{i}(\g(\sigma_{i,i}+T_{i})))$ is in $\partial B(\g_{H},\epsilon_{2})$.
 
To finish the proof of the persistence of the geometrization one still needs to show that the compliment of the persistent 
pieces $(H_{i},\g_{H,i})$ is the $G$ sector or in other words that for any $\epsilon>0$, $(\Sigma^{\epsilon}(\sigma),\g(\sigma))$ converges to 
the $\epsilon$-thick part of the persistent pieces 
$(H_{i},\g_{H,i})$. The proof of this fact follows by contradiction. If this is not the case one can extract a DSLT containing 
an $H$ piece different  from the pieces $(H_{i},\g_{H,i})$. One can prove again that this new piece is persistent leading into 
a contradiction for if persistent, the piece must be one of the pieces $(H_{i},\g_{H,i})$  by the way these pieces 
are defined.  
\ep

\begin{T} Say $Y(\Sigma)\leq 0$ and say $(\g,\K)$ is cosmologically normalized flow satisfying the curvature
assumption. Then $\lim_{\epsilon\rightarrow 0}(\lim_{\sigma} {\mathcal{V}}(\Sigma_{\epsilon}))=0$.
\end{T}

\n {\bf Proof:}

First observe that as the geometrization is persistent and the reduced volume is monotonic, each one of the limits, in
$\sigma$ first and in $\epsilon$ later exists. 
     
Consider the sets $\Omega_{H,\Gamma}(\sigma)=\{x\in \Sigma / |\tilde{\hat{Ric}}(x,\sigma)|_{\g}\leq \Gamma\}$. These sets have 
ATRV when $\sigma\rightarrow \infty$ for any fixed $\Gamma$ and $r$. 
Now consider the set $\Omega_{H,\Gamma,r}(\sigma)=\{x\in\Omega_{H,\Gamma}/|\Rt(x',\sigma)|_{\g}\leq 2\Gamma,\ 
{\rm for\ all}\ x'\in B(x,r)\}$. 
This set has ATRV because the complement is contained in the set $B(\Omega_{\tilde{\hat{Ric}},\Gamma}(\sigma),r)$ which we know must have AZRV
by Propositions \ref{P2} and \ref{P3}. We will need an isoperimetric inequality for the balls of radius one of the regions $\Sigma_{\epsilon}\cap \Omega_{H,\Gamma, r}$ for
a suitable value of $r$ and $\Gamma$. These values of $r$ and $\Gamma$ will come out later using the proposition below. 
Recall Margulis lemma (\cite{Th} corollary 5.10.2)

\begin{Lem} \label{Mar} (Margulis) There is $\epsilon_{0}$ such that for any complete hyperbolic three-manifold $\Sigma$, $\Sigma_{\epsilon_{0}}$ is
(the $\Sigma_{\epsilon_{0}}$ part of) one of the following models:

\begin{enumerate}

\item A horoball modulo $\field{Z}$ or $\field{Z}\times \field{Z}$ (where the action on the half-space model is by horizontal Euclidean translations) or, 

\item a ball around a geodesic $\gamma$ of some radius $R$ modulo $\field{Z}$ (where the action is by translations along the geodesic $\gamma$).

\end{enumerate}

\end{Lem}

We use such $\epsilon_{0}$ in the proposition below.

\begin{Prop}\label{PM} For any $\delta>0$, there is $\epsilon,\sigma_{0},r>1$ and $\Gamma$ such that at any $\sigma\geq \sigma_{0}$, and
for any ball $B(x,r)$ with $x\in\Sigma_{\epsilon}(\sigma)\cap\Omega_{H,\Gamma,r}(\sigma)$ we can unwrap the ball to have
$inj(x)\sim \epsilon_{0}$ and such that on it the ball of radius one is $\delta$-close in $C^{2,\beta}$ to a ball of
radius one in one of the Margulis models 
above.
\end{Prop}

\n{\bf Proof} (of Proposition \ref{PM}): 

Suppose by contrary there is a $\delta>0$ such that the proposition doesn't hold. Then for
every $i$ the conclusion is false for the set of parameters $\Gamma=1/i$, $r=i$, $\sigma_{0}=i$ and $\epsilon(i)$ chosen
in such a way that (at any logarithmic time) for any ball $B(x,r=i)$ with 
$x\in \Sigma_{\epsilon(i)}\cap \Omega_{H,\Gamma=1/i,r=i}$ we can unwrap the ball to have $inj_{\g}(x)\sim \epsilon_{0}$. As we are assuming 
the conclusion is false for any $i$, we can find for any $i$, $\sigma_{i}>\sigma_{0}=i$ and $x_{i}$ in $\Sigma_{\epsilon_{i}}(\sigma_{i})\cap 
\Omega_{H,\Gamma=1/i,r=i}(\sigma)$ such the unwrapped ball is $\delta$-far from one of the Margulis models above. Now the
sequence (in $i$) of such unwrapped balls converges in $C^{2,\beta}$ to a complete Riemannian manifold and because $x_{i}$ is 
in $\Omega_{H,\Gamma=1/i,r=i}(\sigma_{i})$ for every $i$ it must be hyperbolic and therefore one of the Margulis models by Lemma \ref{Mar}. This
is a contradiction. \hspace{\stretch{1}}$\Box$\\
 
Now observe that at each point in any of the Margulis models there is one and only one direction where the size of 
the (collapsed) fibers expands most.
In the first two examples the directions are determined by the vertical geodesic congruence and in the third by the 
congruence of geodesics coming
out perpendicularly from the geodesic $\gamma$. Observe that the direction is invariant under wrappings or 
unwrapping. Define in each model a field $X$ having norm one and 
in the direction of maximal fiber expansion. For example if the model is a cusp, i.e. a horoball 
modulo $\field{Z}\times\field{Z}$ then (writing the metric as $g_{H}=dx^{2}+e^{2x}h$ where $h$ is the flat metric
in the two-torus induced by the action of $\field{Z}\times\field{Z}$ in $\field{R}^{2}$) the vector field $X$ is $\partial_{x}$. 
It is a straight forward 
calculation that the divergence of $X$ in any one of the models is positive and bounded below and above say by 
$C_{1}$ and $C_{2}$ ($0<C_{1}<C_{2}$). For example in the cusp case the divergence of $X$ is computed as $\nabla. X=\frac{1}{\sqrt{|g_{H}}|}\partial_{x}(\sqrt{|g_{H}|})=2$
with $|g_{H}|=A^{2}e^{4x}$ the determinant of the metric $g_{H}$ in the coordinates $(x,\theta_{1},\theta_{2})$ ($(\theta_{1}$ and $\theta_{2}$ are the natural
coordinates on $T^{2}=S^{1}\times S^{1}$ and A is the area under $h$). Suppose now that
we have a manifold $U$ made out of a finite (but arbitrary) set of balls of radius one taken from any one of the 
three Margulis models. The balls may touch each other in an arbitrary fashion, and therefore the boundary may not be 
entirely smooth although it is in a set of total measure. Then
\ben\label{II}
C_{1}Vol(U)\leq \int_{U} \nabla.X dv = \int_{\partial(U)} <X,n>dS \leq Vol(\partial U),
\een

\noindent (where we have used the fact that $X$ is of norm one and therefore $<X,n>\leq 1$). 
Observe also that for any $s<1$ 
\begin{equation}\label{II2}
Vol(B(\partial U,s))\geq C(s) Vol(U).
\end{equation}

\n To see that observe that every point in the 
smooth part of the boundary
joins with one and only one closest center. Then using the isoperimetric inequality \ref{II} above, the set formed 
by all segments of length $s$ 
starting at the points in the smooth parts of the boundary and in the direction of their unique center must have a 
definite part of 
the total volume. We have 
similar inequalities if the balls are made out of balls $\delta$-close in
$C^{2,\beta}$ to one of the Margulis models for $\delta$ sufficiently small. From now on take such $\delta$ in 
Proposition \ref{PM}, 
to get the parameters $\Gamma$, $r$ and $\epsilon$.  

Now lets go back to finish the proof of Theorem \thetwo. Assume by the contrary there is a sequence 
$\{\sigma_{i}\}$ such that $\lim_{\bar{\epsilon}\rightarrow 0}(\lim_{\sigma_{i}\rightarrow \infty} 
{\mathcal{V}}(\Sigma_{\bar{\epsilon}}))={\mathcal{V}}_{0}>0$. We then have $\lim_{\bar{\epsilon}\rightarrow 0}
(\lim_{\sigma_{i}
\rightarrow \infty}{\mathcal{V}}(\Sigma_{\bar{\epsilon}}\cap \Omega_{H,\Gamma,r}))={\mathcal{V}}_{0}$. Fix
$\bar{\epsilon}\leq \epsilon$. We can use the isoperimetric inequality (\ref{II2}) with $s=1/2$, to conclude that 
$\lim_{\sigma_{i}\rightarrow \infty}{\mathcal{V}}(B(\partial B(\Omega_{H,\Gamma,r}(\sigma_{i})\cap \Sigma_{\bar{\epsilon}}(\sigma_{i}),1),s))$ 
is bounded below by a nonzero constant independent of $\bar{\epsilon}$. Note that the set $B(\partial B(\Omega_{H,\Gamma,r}(\sigma_{i})\cap \Sigma_{\bar{\epsilon}}(\sigma_{i}),1),s))$ 
is disjoint from the set $\Omega_{H,\Gamma,r}(\sigma_{i})\cap \Sigma_{\bar{\epsilon}}(\sigma_{i})$ and that the $\sup\{inj(x)_{\g}/ x\in B(\partial B(\Omega_{H,\Gamma,r}(\sigma_{i})\cap \Sigma_{\bar{\epsilon}}(\sigma_{i}),1),s))\}$
converges to zero\footnote{Under curvature bounds the injectivity radius propagates over finite distances, in particular if $inj_{\g}(x)\rightarrow 0$ and $d(x,y)<C$ then $inj_{\g}(y)\rightarrow 0$.} as $i\rightarrow \infty$, therefore $\lim_{\bar{\epsilon}}(\lim_{\sigma\rightarrow \infty}{\mathcal{V}}(\Sigma_{\bar{\epsilon}}))>{\mathcal{V}}_{0}$
which is a contradiction \hspace{\stretch{1}}$\Box$\\

We prove next Corollaries \ref{cor1}-\ref{cor5}. Corollary $2$ is direct from Theorem \thetwo. For corollary $3$ observe that
since ${\mathcal{V}}$ is monotonic there cannot be any $H$ piece emerging and so we are in the situation of Corollary $2$. Corollary $5$
is the content of Proposition \ref{P3} {\it 5}. To prove Corollary $1$ observe that as proved in \cite{R} (Theorem 9) given $\Lambda$
there is $\epsilon$ such that if ${\mathcal{V}}(\g,\K)-{\mathcal{V}}_{inf}<\epsilon$ then the thick-thin decomposition
implements the Thurston geometrization, therefore if the flow starts in a state $(\g,\K)(\sigma_{0})$ with ${\mathcal{V}}(\g,\K)(\sigma_{0})-
{\mathcal{V}}_{inf}<\epsilon$ as ${\mathcal{V}}$ is monotonic the difference ${\mathcal{V}}(\g,\K)(\sigma)-{\mathcal{V}}_{inf}$ 
is kept along the evolution, so the Thurston geometrization is the persistent geometrization. By Theorem \thetwo
it must be ${\mathcal{V}}\downarrow {\mathcal{V}}_{inf}$. For Corollary \ref{cor4} observe again that by what has been proved
in \cite{R} and Theorems $1$ and $2$, ${\mathcal{V}}\downarrow {\mathcal{V}}_{inf}$ iff the persistent long time geometrization is the Thurston geometrization
iff the tori separating the $H$ and $G$ sectors are incompressible.

\end{document}